\documentclass{article}

\usepackage{arxiv}

\usepackage[utf8]{inputenc} % allow utf-8 input
\usepackage[T1]{fontenc}    % use 8-bit T1 fonts
\usepackage{hyperref}       % hyperlinks
\usepackage{url}            % simple URL typesetting
\usepackage{booktabs}       % professional-quality tables
\usepackage{amsfonts}       % blackboard math symbols
\usepackage{nicefrac}       % compact symbols for 1/2, etc.
\usepackage{microtype}      % microtypography
\usepackage{lipsum}		% Can be removed after putting your text content
\usepackage{float} % For [H] float specifier
\usepackage{graphicx}
\usepackage{doi}
\usepackage{amsmath}
\usepackage{float}
\usepackage{bbding}
\usepackage{pifont}
\usepackage{wasysym}
\usepackage{amssymb}
\usepackage{xcolor}
\usepackage{titling}
\usepackage{lmodern}

\pretitle{%
  \par\vspace{-1em}
  \hrule height 1.2pt % bold top rule
  \vspace{0.8em}
  \begin{center}\LARGE\bfseries
}
\posttitle{%
  \par\end{center}
  \vspace{0.8em}
  \hrule height 1.2pt % bold bottom rule
  \vspace{1em}
}

\preauthor{\begin{center}}
\postauthor{\end{center}}
\predate{\begin{center}\small\itshape}
\postdate{\end{center}\vspace{-1em}}

\definecolor{darkgreen}{RGB}{0,100,0} % Define dark green color

\newcommand{\cmark}{\textcolor{darkgreen}{\checkmark}} % Green checkmark
\newcommand{\xmark}{\textcolor{red}{$\times$}} % Red cross mark

\title{PINNs for Solving Unsteady Maxwell's Equations: Convergence Issues and Comparative Assessment with Compact Schemes}  
\author{
  Gal G. Shaviner$^{\text{a}}$ \quad 
  Hemanth Chandravamsi$^{\text{a}}$ \quad
  Shimon Pisnoy$^{\text{a}}$ \quad
  Ziv Chen$^{\text{b}}$ \quad
  Steven H. Frankel$^{\text{a}}$
  \vspace{5pt} \\
  $^{\text{a}}$Faculty of Mechanical Engineering, Technion - Israel Institute of Technology, Haifa, Israel 3200003 \\
  $^{\text{b}}$Faculty of Electrical and Computer Engineering, Technion - Israel Institute of Technology, Haifa, Israel 3200003 \\
  \noindent \texttt{\{gal.shaviner, hemanth, shimonpi, ziv.chen\}@campus.technion.ac.il}, \texttt{frankel@me.technion.ac.il}
}

\begin{document}
\maketitle
\begin{abstract}

Physics-Informed Neural Networks (PINNs) have recently emerged as a promising alternative for solving partial differential equations, offering a mesh-free framework that incorporates physical laws directly into the learning process. 
In this study, we explore the application of Physics-Informed Neural Networks (PINNs) for solving unsteady Maxwell’s equations and compare their performance with two established numerical methods: the Finite-Difference Time-Domain (FDTD) method and a compact Pade scheme with filtering. Three benchmark problems are considered, ranging from 1-D free-space wave propagation to 2-D Gaussian pulses in periodic and dielectric media. We assess the effectiveness of convergence-enhancing strategies for PINNs, including random Fourier features, spatio-temporal periodicity, and temporal causality training. An ablation study highlights that architectural choices must align with the underlying physics. Additionally, we employ a Neural Tangent Kernel framework to examine the spatio-temporal convergence behavior of PINNs. Results show that convergence rates correlate with error over time but not in space, revealing a limitation in how training dynamics allocate learning effort. Overall, this study demonstrates that PINNs, when properly configured, can match or surpass traditional solvers in accuracy and flexibility, though challenges remain in addressing spatial inhomogeneity and adapting training to localized complexity.

%Physics-Informed Neural Networks (PINNs) have recently emerged as a promising alternative for solving partial differential equations, offering a mesh-free framework that incorporates physical laws directly into the learning process. In this study, we explore the capabilities of PINNs in solving one- and two-dimensional Maxwell’s equations and show that, with the use of convergence-enhancing techniques, PINNs can accurately capture complex wave phenomena—including sharp gradients and broadband transients—that pose challenges for traditional methods. We compare the performance of PINNs against Finite-Difference Time-Domain (FDTD) and high-order compact finite-difference schemes. While these traditional approaches remain effective in many settings, they exhibit limitations near discontinuities and in resolving high-frequency content. Our results demonstrate that PINNs not only match but in some cases surpass these methods in both accuracy and robustness. This work positions PINNs as a viable tool for solving Maxwell’s equations, particularly in scenarios where standard approaches fall short.
\end{abstract}

% keywords can be removed

\section{Introduction}
% Physics-Informed Neural Networks (PINNs) have recently gained substantial attention for their ability to solve ordinary differential equations (ODEs) and partial differential equations (PDEs) across a wide range of scientific and engineering applications.

% Maxwell Introduction
Maxwell's equations, together with the Lorentz force law, form the foundation of classical electromagnetism and underpin a wide range of applications in optics, wireless communications, and electrical circuits \cite{lam2003wireless, makarenko2020generalized, thompson2012maxwell}. These equations are non-linear and hyperbolic, making them difficult to solve analytically, especially in higher dimensions. Consequently, numerical methods such as the finite-difference time-domain (FDTD) method \cite{JENKINSON2018192}, finite-element methods \cite{CHANG198889}, and finite-volume methods \cite{HERMELINE20089365} are widely used to calculate approximate solutions. Although effective, these methods suffer from computational costs that scale exponentially with dimensionality, typically as \( \mathcal{O}(N^d) \), a manifestation of the curse of dimensionality. Additionally, traditional solvers often face stability constraints, limited flexibility in handling complex geometries due to mesh requirements, and challenges in incorporating sparse or noisy experimental data.

% PINNs introduction - upsides and downsides?
Physics-Informed Neural Networks (PINNs) \cite{raissi2019physics} have emerged as a promising alternative for solving ordinary and partial differential equations with little to no training data. Unlike traditional numerical methods, PINNs provide a mesh-free framework that learns implicit solutions by embedding physical laws directly into the optimization process. Despite being mathematically well-posed, PINNs often suffer from slow convergence and sensitivity to hyperparameters. These difficulties are largely attributed to spectral bias \cite{tancik2020, wang2020eigenvector} and imbalanced gradient magnitudes between different loss components during training \cite{wang2022and}. Even with recent advances in network architectures and training strategies, accurately capturing high-frequency features, such as turbulent structures or broadband signals, remains a significant challenge. As a result, traditional numerical solvers continue to be more practical in many real-world applications.

% optimism - solutions to pinns challenges
Recent advances in optimization strategies \cite{wang2023experts, toscano2025pinns} have led to notable improvements in PINNs performance.  As research continues to address training pathologies and convergence issues, PINNs are becoming increasingly viable for solving complex electromagnetic problems. Several key strategies have been proposed to enhance convergence and stability:

\begin{itemize}
    \setlength{\itemsep}{0pt}
    \setlength{\parskip}{0pt}
        \item \textbf{Random Fourier Features (RFF) and variants:} To address spectral bias—where neural networks tend to prioritize low-frequency solutions \cite{rahaman2019, basri2020}—RFF embeddings transform input coordinates into high-frequency signals before passing them through the network \cite{rahimi2007random, tancik2020}. This improves the network’s ability to capture fine-scale features. Wang et al. \cite{wang2020eigenvector} further developed a multi-scale Fourier embedding to address multiscale spatio-temporal problems.
        \item \textbf{Random Weight Factorization (RWF):} Introduced by Wang et al. \cite{wang2022random}, RWF improves training stability and accuracy by restructuring the network’s parameterization to enhance gradient flow and representation.
        \item \textbf{Strict Boundary Condition Enforcement:} Dong et al. \cite{dong2021} proposed a method for enforcing periodic boundary conditions as hard constraints, eliminating the need to include them in the loss function and thereby improving convergence.
        \item \textbf{Adaptive Temporal Weighting:} To improve learning in time-dependent problems, Wang et al. \cite{wang2022} proposed partitioning the temporal domain into \(M\) segments, each with its own weight. This prevents the network from focusing prematurely on later time steps and ensures better resolution of early-time dynamics.
        \item \textbf{Second-order optimizers:} First-order optimizers like Adam and SGD often fail to efficiently navigate the complex loss landscapes of PINNs. Second-order methods such as L-BFGS \cite{liu1989limited}, NysNewton \cite{rathore2024challenges, gu2023nysnewton}, and self-scaled BFGS (SS-BFGS) \cite{oren1974self, kiyani2025optimizer} leverage curvature information to accelerate convergence and improve training robustness.
\end{itemize}

Additional strategies that have been proposed to improve the training and performance of PINNs include residual-based attention mechanisms \cite{anagnostopoulos2024residual}, loss balancing techniques \cite{wang2021understanding, xiang2022self, mcclenny2023self, wang2022and}, adaptive sampling methods \cite{wu2023comprehensive, lin2024causality}, domain decomposition approaches \cite{jagtap2020conservative, wight2020solving}, and curriculum learning strategies \cite{krishnapriyan2021characterizing}. Together, these developments have enabled PINNs to be rapidly adopted across a wide range of scientific disciplines, including robotics \cite{liu2024physics}, control systems \cite{Gu2024}, computer vision \cite{Liu2020}, fluid dynamics \cite{Guo2016, Carlberg2019, Safikhani2011}, agriculture \cite{Kamilaris2018}, and electromagnetism \cite{Zheng2024}.

Within the domain of electromagnetism, PINNs have been applied to various formulations of Maxwell’s equations. However, their ability to reliably capture time-dependent electromagnetic phenomena remains insufficiently explored. In this work, we investigate the effectiveness of convergence-enhancing strategies for training PINNs on unsteady Maxwell’s equations in simple 2-D rectangular domains.

% pinns for maxwell - whats there in the lit - whats missing?
Several studies have explored the application of PINNs to Maxwell’s equations. Zheng et al. \cite{Zheng2024} investigated the role of network depth in solving 1-D Maxwell problems. Nohra and Dufour \cite{Nohra2024} applied PINNs to steady-state 1-D, 2-D, and 3-D Maxwell equations, and analyzed the training dynamics of first- and second-order formulations using Neural Tangent Kernel (NTK) theory \cite{jacot2018neural}, showing that first-order formulations are particularly effective in handling discontinuities and sharp gradients. Zhang and Fu \cite{Zhang2022} developed a physics-informed deep neural network for Maxwell’s plasma coupling system, enabling global field reconstruction and inversion of inhomogeneous plasma parameters, validated through 1-D wave propagation in magnetized plasmas. Additional contributions on PINNs applied to Maxwell’s equations include works on magnetostatics, inverse problems, and quasi-static formulations \cite{chen2021physics, kovacs2021magnetostatics, piao2023domain, brendel2022physics, lim2022maxwellnet}.

While deep neural networks used in PINNs are highly expressive, obtaining accurate solutions remains challenging due to the soft imposition of physics constraints through the PDE loss term. As noted by Krishnapriyan et al. \cite{krishnapriyan2021characterizing}, this formulation often leads to convergence issues, particularly in the presence of complex dynamics. Importantly, there remains limited investigation into the training challenges associated with unsteady two-dimensional Maxwell’s equations. Motivated by this gap, the present study investigates strategies to improve PINN training for such problems and benchmarks their performance against conventional numerical solvers.

This study aims to address the following key questions:

\begin{enumerate} 
    \item What enhancements beyond the baseline PINN model of Raissi et al. ~\cite{raissi2019physics} are necessary to accurately and reliably solve the 2-D unsteady Maxwell equations? 
    \item How do PINNs compare with traditional numerical methods in terms of solution accuracy and computational efficiency? 
    \item Do PINNs naturally allocate training effort by converging more rapidly in regions with larger PDE residuals, while under-emphasizing areas with smaller errors? 
\end{enumerate}

% We answer this by performing an ablation study employing different enhancements including Random Fourier Features, Strict periodic conditions, and training that respects the principles of temporal causality. 

% aim/outline of the paper (optional)
% We solve Maxwell's equations both in one and two dimensions in the TM (transverse magnetic) mode. We implement RFF and RWF in a neural network structure, furthermore we apply strict periodic boundary conditions. In addition we assign the PDEs loss function temporal weights in order to respect the temporal causality. We solve Maxwell's equations with high order numerical schemes in order to compare our results to PINNs.

\section{Maxwell's equations}
The microscopic form of Maxwell’s equations governs the behavior of electric and magnetic fields in space and time. They are expressed as:

\begin{equation}
\nabla \cdot \mathbf{E} = \frac{\rho}{\varepsilon}  
\end{equation} \label{eq:maxwell1} 
\begin{equation}
\nabla \cdot \mathbf{B} = 0 \end{equation} \label{eq:maxwell2}
\begin{equation}
\nabla \times \mathbf{E} = - \frac{\partial \mathbf{B}}{\partial t} \end{equation} \label{eq:maxwell3} 
\begin{equation}
\nabla \times \mathbf{B} = \mu \left(\mathbf{J} + \varepsilon \frac{\partial \mathbf{E}}{\partial t}\right) \end{equation} \label{eq:maxwell4}

Here $\mathbf{E}$ and $\mathbf{B}$ denote the electric and  magnetic fields; $\rho$ the electric charge density; $\mathbf{J}$ is the current density; $\varepsilon$ is the electric permittivity; and $\mu$ is the magnetic permeability.

In regions free of charges and currents, these equations simplify to:
\begin{align}
    \nabla \times \mathbf{B} &= \mu \varepsilon \frac{\partial \mathbf{E}}{\partial t}\label{eq:maxwell_simplified1} \\
    \nabla \times \mathbf{E} &= -\frac{\partial \mathbf{B}}{\partial t}\label{eq:maxwell_simplified2}
\end{align}
For linear, isotropic, and homogeneous materials, the constitutive relations connecting the electric field $\mathbf{E}$, magnetic field $\mathbf{B}$, electric displacement field $\mathbf{D}$, and the magnetizing field $\mathbf{H}$ are given by:
\begin{equation}
    \mathbf{D} = \varepsilon \mathbf{E}
\end{equation}
\begin{equation}
    \mathbf{B} = \mu \mathbf{H} \label{eq:BmuH}
\end{equation}
where $\varepsilon$ and $\mu$ denote the permittivity and permeability of the medium, respectively.

Substituting Eq. ~\ref{eq:BmuH} into Eq.~\ref{eq:maxwell_simplified1} and Eq. \ref{eq:maxwell_simplified2}, we obtain the Maxwell equations in terms of the electric field $\mathbf{E}$ and the magnetizing field $\mathbf{H}$:

\begin{align}
    \frac{\partial \mathbf{E}}{\partial t} &= \frac{1}{\varepsilon} \nabla \times \mathbf{H}\label{eq:maxwell_eh1} \\
    \frac{\partial \mathbf{H}}{\partial t} &= -\frac{1}{\mu} \nabla \times \mathbf{E}\label{eq:maxwell_eh2}
\end{align}

For wave propagation in one spatial dimension along the $z$-axis, Maxwell’s equations reduce to:

\begin{align} 
    \frac{\partial E_x}{\partial t} &= -\frac{1}{\varepsilon} \frac{\partial H_y}{\partial z}\label{eq:maxwell_1d_1} \\
    \frac{\partial H_y}{\partial t} &= -\frac{1}{\mu} \frac{\partial E_x}{\partial z}\label{eq:maxwell_1d_2}
\end{align}

In two spatial dimensions, Maxwell’s equations take the form:

\begin{align} 
    \frac{\partial E_z}{\partial t} &= \frac{1}{\varepsilon} \left( \frac{\partial H_y}{\partial x} - \frac{\partial H_x}{\partial y} \right) \label{eq:maxwell_2d_1} \\
    \frac{\partial H_x}{\partial t} &= -\frac{1}{\mu} \frac{\partial E_z}{\partial y}\label{eq:maxwell_2d_2} \\
    \frac{\partial H_y}{\partial t} &= \frac{1}{\mu} \frac{\partial E_z}{\partial x}\label{eq:maxwell_2d_3}
\end{align}  \label{eqn:maxwell2D}

For free-space (vacuum), the material parameters are normalized such that $\varepsilon = \mu = 1$.

\section{Model and training}
\subsection{Network architecture}
A fully connected feedforward neural network is employed to approximate the solution of Maxwell’s equations by mapping the input vector \(\mathbf{x} = (x, y, t)\), representing spatial and temporal coordinates, to the output vector \(\mathbf{u} = (E_z, H_x, H_y)\), corresponding to the field components in two dimensions. 

For problems involving spatial periodicity (see case 2 in section \ref{sec:expts}), periodic boundary conditions are enforced explicitly by mapping the original input \( \mathbf{x}: \mathbb{R}^3 \) to a periodic coordinate representation \( \tilde{\mathbf{x}}: \mathbb{R}^6 \), following the method proposed by Dong et al. \cite{dong2021}:

\begin{equation}
    \tilde{\mathbf{x}} = \mathcal{P}(\mathbf{x}) = \left[\sin \left(\frac{2 \pi}{L_x} x\right), \cos \left(\frac{2 \pi}{L_x} x\right), \sin \left(\frac{2 \pi}{L_y} y\right), \cos \left(\frac{2 \pi}{L_y} y\right), \sin \left(\frac{2 \pi}{P_t} t\right), \cos \left(\frac{2 \pi}{P_t} t\right) \right]
\end{equation}

Here, $L_x$ and $L_y$ denote the lengths in the $x$ and $y$ directions, respectively. Although the problems considered are not inherently periodic in time, we introduce a trainable temporal period \( P_t \), which was found to improve convergence, consistent with observations in prior studies \cite{wang2023experts}.

To address spectral bias—where neural networks preferentially learn low-frequency components \cite{basri2020}, the periodic coordinates $\tilde{\mathbf{x}}$ are further mapped to a higher-dimensional space using RFF ~\cite{rahimi2007random, tancik2020}:

\begin{equation}
\Phi(\tilde{\mathbf{x}}) = \left[\sin(\mathbf{B} \tilde{\mathbf{x}}), \cos(\mathbf{B} \tilde{\mathbf{x}}) \right]
\end{equation}

where \( \mathbf{B} \in \mathbb{R}^{H \times 6} \) is a fixed random matrix with entries sampled from a normal distribution with mean \( \mu = 0 \) and standard deviation \( \sigma = 2 \), and \( H \) is the width of the first hidden layer.

The neural network weights are initialized using the RWF technique proposed by Wang et al.~\cite{wang2022random}. The network architecture is defined as follows:

\begin{equation}
\mathbf{h}^{(0)} = \sigma(\mathbf{W}^{(0)} \Phi(\mathcal{P}(\mathbf{x})) + \mathbf{b}^{(0)}) , \quad \text{(input to first hidden layer)}
\end{equation}
\begin{equation}
\mathbf{h}^{(l)} = \sigma(\mathbf{W}^{(l)} \mathbf{h}^{(l-1)} + \mathbf{b}^{(l)}), \quad l = 1, 2, \dots, L \quad \text{(first to last hidden layer)}
\end{equation}
\begin{equation}
\mathbf{u} = \mathbf{W}^{(L+1)} \mathbf{h}^{(L)} + \mathbf{b}^{(L+1)}, \quad \text{(last hidden layer to output layer)}
\end{equation}

\begin{figure}[H]
    \centering
    \includegraphics[width=1.0\textwidth]{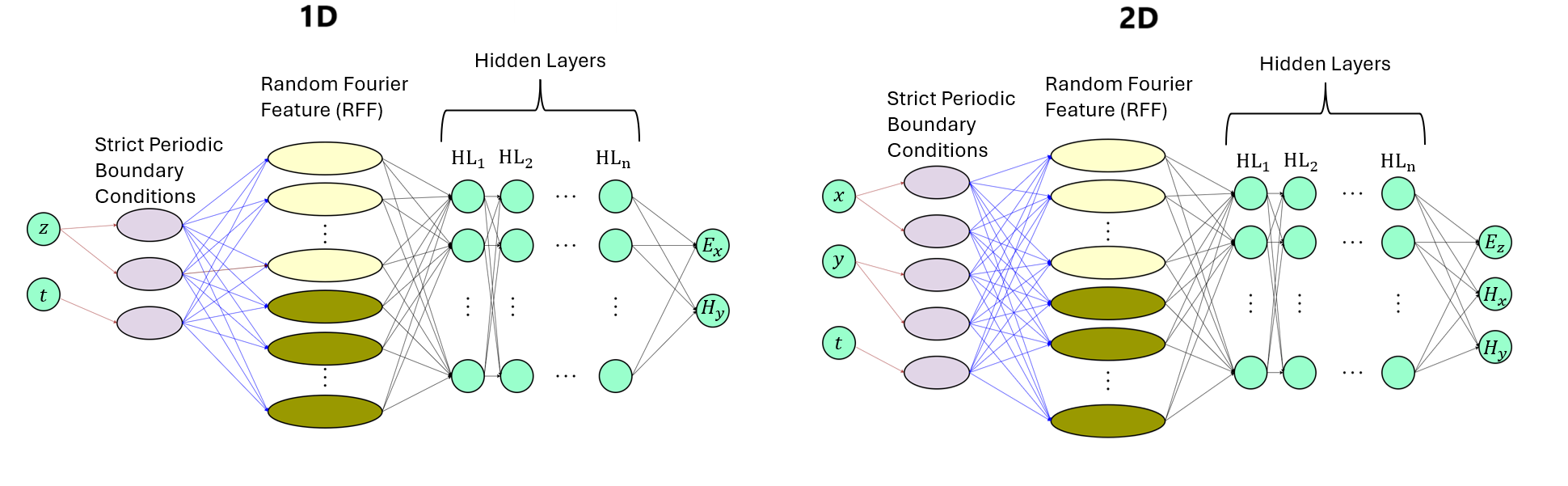}
    \caption{Schematic of PINN architectures for solving Maxwell's equations in 1D (left) and 2D (right). In the case of 1D, the network takes \( z \) and \( t \) as inputs and predicts \( E_x \) and \( H_y \). For the 2D cases, the inputs \( (x, y, t) \) map to outputs \( (E_z, H_x, H_y) \). A periodic mapping layer first enforces strict periodic boundary condition constraints (in space only), followed by a Random Fourier Features (RFF) layer that maps \( \tilde{\mathbf{x}} \) to a \( 2H \)-dimensional space. The transformed inputs then pass through the hidden layers, before producing the final outputs.}
    \label{fig:architecture}
\end{figure}

Here, $\sigma$ is the non-linear activation function for which the hyperbolic tangent ($\tanh$) is used. A schematic illustration of network architecture for both 1-D and 2-D Maxwell's equations is provided in Figure \ref{fig:architecture}.

\subsection{Loss function}
The loss function used to train the PINN comprises three components: the initial condition loss $\mathcal{L}_{\text{IC}}$, the boundary condition loss $\mathcal{L}_{\text{BC}}$, and the residual loss $\mathcal{L}_{\text{res}}$, which enforces Maxwell's equations. For problems with periodic boundary conditions, the boundary loss term is omitted, as the periodic mapping inherently satisfies these constraints. The total loss over the spatio-temporal domain $\Omega \times [0,T]$ is defined as:
\begin{equation}
    \mathcal{L}(\mathbf{x}, t) = \lambda_{\text{IC}} \mathcal{L}_{\text{IC}} + \lambda_{\text{BC}} \mathcal{L}_{\text{BC}} + \lambda_{\text{res}} \mathcal{L}_{\text{res}}
    \label{eq:total_loss}
\end{equation}

where $[\lambda_{\text{IC}}, \lambda_{\text{BC}}, \lambda_{\text{res}}]$ are coefficients that balance the contribution of each loss term. For the 1-D case, $\mathbf{x} = z$; for 2-D, $\mathbf{x}=(x,y)$. 

The individual components of the loss are defined as follows:
\begin{equation}
    \mathcal{L}_{\text{IC}}(\mathbf{x},t=0) = \frac{1}{N_{\text{IC}}} \sum_{i=1}^{N_{\text{IC}}} \left\| \mathbf{u}(\mathbf{x}_{\text{IC}}^i, t=0) - \mathbf{u}_{\text{exact}}(\mathbf{x}_{\text{IC}}^i, t=0) \right\|^2 
\end{equation}
where $\,\,\mathbf{x}_{\text{IC}} =( x_{\text{IC}}, y_{\text{IC}}, z_{\text{IC}}) \in \{(x, y, z) \mid t = 0\}$, and $N_{\text{IC}}$ is the number of interior spatial points used to enforce the initial condition.
\begin{equation}
    \mathcal{L}_{\text{BC}}(\mathbf{x}, t) = \frac{1}{N_t} \frac{1}{N_{\text{b}}} \sum_{j=0}^{N_t} \sum_{i=1}^{N_{\text{b}} } \left\| \mathcal{B} [\mathbf{u}] (\mathbf{x}_{\text{b}}^i, t^j)  \right\|^2
\end{equation}
Where $\,\,\mathbf{x}_{\text{b}} = (x_{\text{b}}, y_{\text{b}}, z_{\text{b}}) \in \partial \Omega\,\,\, \text{and} \,\,\,t^j \in t$, $N_{\text{b}}$ is the number of boundary points, and $N_{\text{t}}$ is the number of sampled time points.
\begin{equation}
    \mathcal{L}_{\text{res}}(\mathbf{x},t) = \frac{1}{N_{\text{t}}}\frac{1}{N_{\text{r}}} \sum_{j=0}^{N_t}\sum_{i=1}^{N_{\text{r}} } \left\| \mathcal{F}(\mathbf{u}, \mathbf{x}_{\text{r}}^i, t^j) \right\|^2
    \label{eq:residual_1d_loss}
\end{equation}
where, $\mathbf{x}_{\text{r}}$ are collocation points sampled within the domain $\Omega$, and $N_{\text{r}}$ is the number of residual points used at each time step. The operator $\mathcal{F}(\mathbf{u}, x, y, z, t)$ is defined as the mismatch between the temporal derivative and the spatial operator from Maxwell’s equations:
\begin{equation}
     \mathcal{F}(\mathbf{u}, \mathbf{x}, t) = 
\frac{\partial \mathbf{u}}{\partial t}\left(\mathbf{x}, t\right)+\mathcal{N}\left[\mathbf{u}\right]\left(\mathbf{x}, t\right)
\label{eq:F_1d}
\end{equation}
where $\mathcal{N}[\mathbf{u}]$ denotes the spatial differential operator that encapsulates the curl terms specific to Maxwell’s equations. All derivatives are computed using automatic differentiation.

Training is performed using the Adam optimizer with an initial learning rate $\eta = 10^{-3}$. A learning rate scheduler is applied, reducing the learning rate by a factor of 0.85 every 2000 epochs across all experiments.

\subsection{Training with temporal causality enforcement}
To mitigate training pathologies in PINNs arising from violations of temporal causality, as identified in \cite{krishnapriyan2021characterizing, wang2022}, we adopt the temporal causality training strategy proposed by Wang et al.~\cite{wang2022}. This approach ensures that the network first learns dynamics at earlier time steps before progressively incorporating later ones, thereby preventing premature overfitting to inaccurate late-time solutions. Following the formulation in \cite{wang2022}, the residual loss defined in Eq.~\ref{eq:residual_1d_loss} is replaced with a weighted temporal loss of the form:
\begin{equation}
    \mathcal{L}(\theta, t) = \frac{1}{N_t} \sum_{i=0}^{N_t} w_i \mathcal{L}(t_i, \theta)
\end{equation}
where $\theta$ denotes the trainable network parameters and $w_i$ is a time-dependent weight assigned to each time instance $t_i$. The weights $w_i$ are updated iteratively to prioritize earlier time steps based on the cumulative loss up to time $t_i$. Specifically, after each training epoch, the weights are updated as:
\begin{equation}
    w_i = \exp \left(-\epsilon \sum_{k=1}^{i-1} \mathcal{L}(t_k, \theta) \right), \quad i = 2, 3, \dots, N_t, \quad (\text{note: } w_0=1)
\end{equation}
Here, $\epsilon > 0$ is a hyperparameter that controls the rate at which later time steps are incorporated into training. A small $\epsilon$ slows the inclusion of future times, allowing the network to focus more effectively on early-time dynamics.

% Using the Neural Tangent Kernel based analysis shown by Wang et al. \cite{wang2022}, we also study if the convergence rate is at all influenced by the temporal causality enforcement.

\section{Results and Analysis} \label{sec:expts}

\subsection{1-D Gaussian pulse in free space} \label{sec:case1}
To examine the differences in training dynamics between 1- and 2-D unsteady problems, we begin with 1-D case governed by Eq. \ref{eq:maxwell_1d_1} and \ref{eq:maxwell_1d_2}. Specifically, we consider the propagation of a time-decaying Gaussian electric pulse in free space (vacuum), following the setup in \cite{Houle}. 

The electric field pulse is imposed as a Dirichlet boundary condition at the left boundary $z=0$:
\begin{equation}
    {E_{x}}(z=0,t) = e^{-0.5(t^2/0.3^2)}
    \label{eqn:GaussianBC}
\end{equation} 
For the magnetic field $H_y$, a Neumann boundary condition is applied at $z=0$ by setting its spatial gradient to zero. At the right boundary \( z = 1 \), Neumann conditions are applied to both $E_x$ and $H_y$. The time domain spans \( t \in [0,1] \), discretized using \( N_t \) uniformly spaced collocation points. The corresponding boundary condition loss is: 
\begin{equation}
\begin{aligned}
L_{\text{BC}} = & \frac{1}{N_t} \sum_{i=0}^{N_t} \Bigg[ \left( E_x(z=0,t^i) - e^{-0.5((t^i)^2/0.3^2)} \right)^2 + \left( \frac{\partial H_y}{\partial z}(z=0,t^i) \right)^2 \\
& \quad + \left( \frac{\partial E_x}{\partial z}(z=1,t^i) \right)^2 + \left( \frac{\partial H_y}{\partial z}(z=1,t^i) \right)^2 \Bigg]
\end{aligned}
\end{equation}
The initial condition at $t=0$ is defined such that the electric field is set to zero for all $z\in(0,1]$, and the magnetic field is zero for all $z\in(0,1]$. The corresponding initial condition loss is:
\begin{equation}
L_{\text{IC}} = \frac{1}{N_{z_{1}}} \sum_{i=1}^{N_{z_{1}}} \big( E_x(z,t=0) \big)^2 
+ 
\frac{1}{N_{z_{2}}} \sum_{i=1}^{N_{z_{2}}} \big( H_y(z,t=0) \big)^2
\end{equation}
where $N_{z_{1}}$ and $N_{z_{2}}$ denote the number of spatial points sampled in the intervals $z\in(0,1)$ and $z\in[0,1]$, respectively.

The residual loss is computed using the governing equations and the general residual loss definition from Eq. \ref{eq:maxwell_1d_1}, \ref{eq:maxwell_1d_2}, \ref{eq:residual_1d_loss}, and \ref{eq:F_1d}. For the 1-D case, this becomes:
\begin{equation}
    \mathcal{L}_{\text{res}} = \frac{1}{N_{\text{t}}}\frac{1}{N_{\text{xy}}} \sum_{j=0}^{N_t}\sum_{i=1}^{N_{\text{xy}}}  
    \Bigg[ 
    \left(\frac{\partial E_x}{\partial t}(z^i,t^j) +
    \frac{\partial H_y}{\partial z}(z^i,t^j)\right)^2
    +
    \left(\frac{\partial H_y}{\partial t}(z^i,t^j) +
    \frac{\partial E_x}{\partial z}(z^i,t^j)\right)^2
    \Bigg]
\end{equation}

The total loss is then formulated as a weighted combination of all components, as described in Eq. \ref{eq:total_loss}. In this case, the initial and boundary condition losses are each weighted by 20.

The 1-D Gaussian pulse problem presents a significant numerical challenge due to the formation of a sharp discontinuity induced by the imposed boundary and initial conditions. To evaluate the effectiveness of PINNs in this context, we compare its performance against two well-established numerical schemes for solving Maxwell's equations: the finite-difference time-domain (FDTD) method and a high-order Pade scheme with filtering. Further implementation details for both schemes are provided in Appendix~\ref{app:numerics}. Figure~\ref{fig:Ez_1D_comparison} presents a comparison of the electric field at time $t = 0.8$, obtained using the three methods. A discontinuity emerges at $z = 0.8$, which poses a stiff challenge for numerical solvers. The FDTD solution displays significant oscillations near the discontinuity, which propagate upstream and distort the field in regions where the solution should remain smooth. The Pade scheme offers moderate improvement, but residual oscillations remain concentrated around the shock. These numerical artifacts persist even in the reference solution, computed using high-resolution $1000 \times 1000$ FDTD with post-processing filters (depicted in blue), despite extensive tuning of filtering parameters and grid refinement. In contrast, the PINN solution accurately captures the field distribution and sharply resolves the discontinuity without introducing spurious oscillations. This highlights a key advantage of PINNs: by minimizing the governing equations directly and leveraging global solution context, they are able to handle sharp transients without the dispersive and dissipative errors typical of traditional grid-based methods.

\begin{figure}[H]
    \centering    \includegraphics[width=1.0\textwidth]{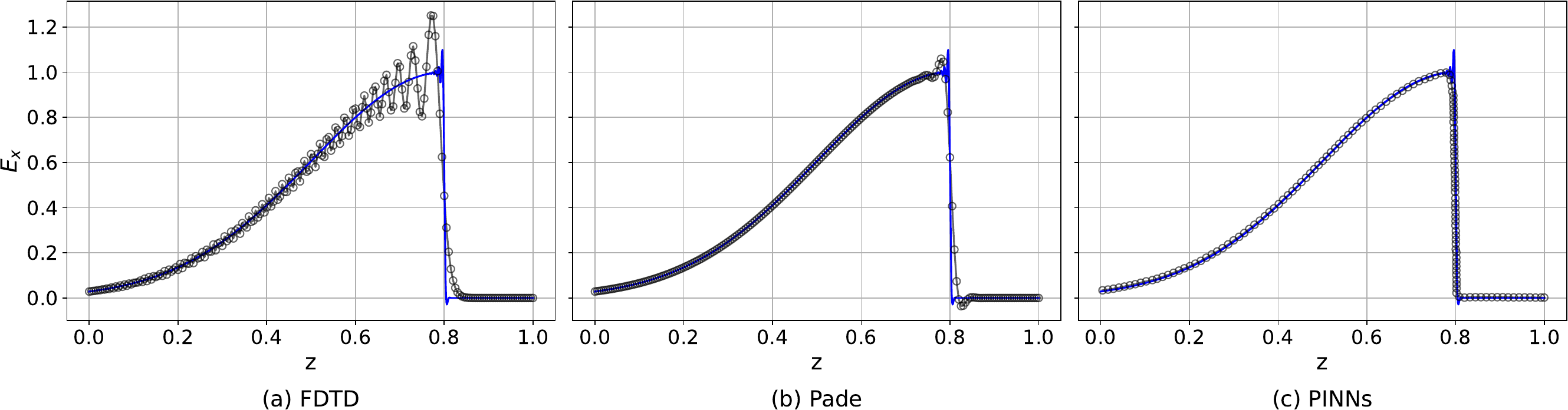}
    \caption{Solution for the electric field  \(E_x\) obtained using (a) FDTD, (b) Pade (4th order), and (c) PINNs at \( t = 0.8 \) for a decaying point valued electric pulse located at \(z=0\). Noticeable oscillations can be seen with FDTD and Pade, particularly near the sharp gradients, while PINNs exhibits an oscillation free and accurate solution. The blue curve indicates the reference solution computed using 1000 spatial points with FDTD and filtering.}
    \label{fig:Ez_1D_comparison}
\end{figure}

\subsection{2-D Gaussian pulse in a periodic box} \label{sec:case2}
We now consider the evolution of a 2-D Gaussian pulse in electric field within a spatially periodic domain. The spatial coordinates are defined as \( x,y \in [-1,1] \), and the temporal domain spans \( t \in [0,1.5] \). When strict periodic boundary conditions are enforced within the neural network architecture, as described by Dong et al.~\cite{dong2021}, an explicit boundary condition loss term is no longer required. The initial condition consists of a Gaussian pulse in the electric field, with zero initial magnetic field. These are specified as follows:

\begin{equation}
{E_{z}}_{0}(x,y,t=0) = e^{-25 (x^2 + y^2)}
\label{eqn:pulse_testcase2_Ez}
\end{equation}
% \vspace{-18mm}
\begin{equation}
{H_{x}}_{0}(x,y,t=0) = 0
\label{eqn:pulse_testcase2_Hx}
\end{equation}
% \vspace{-20mm}
\begin{equation}
{H_{y}}_{0}(x,y,t=0) = 0
\label{eqn:pulse_testcase2_Hy}
\end{equation}

The initial condition loss is given by:

\begin{equation}
L_{\text{IC}} = \frac{1}{N_{xy}} \sum_{i=1}^{N_{xy}} \Bigg[
\big( {E_{z}}_{0}(x^i,y^i,t=0) \big)^2
+ \big( {H_{x}}_{0}(x^i,y^i,t=0) \big)^2
+ \big( {H_{y}}_{0}(x^i,y^i,t=0) \big)^2
\Bigg] 
\label{eqn:2D_IC}
\end{equation}

where $N_{{xy}}$ is the number of spatial points in the 2-D domain.

The residual loss is calculated using to Eq. \ref{eq:maxwell_2d_1}-\ref{eq:maxwell_2d_3}, \ref{eq:residual_1d_loss}, and \ref{eq:F_1d}, as:

\begin{equation}
\begin{aligned}
    \mathcal{L}_{\text{res}} = \frac{1}{N_{\text{t}}}\frac{1}{N_{\text{xy}}} \sum_{j=0}^{N_t}\sum_{i=1}^{N_{\text{xy}}}  
    \Bigg[ 
    & \left(\frac{\partial E_z}{\partial t}(x^i,y^i,t^j) -
    \frac{1}{\varepsilon}\left(\frac{\partial H_y}{\partial x}(x^i,y^i,t^j)-\frac{\partial H_x}{\partial y}(x^i,y^i,t^j)\right)^2\right)
    +
    \\
    & \left(\frac{\partial H_x}{\partial t}(x^i,y^i,t^j) +
    \frac{1}{\mu}\frac{\partial E_z}{\partial y}(x^i,y^i,t^j)\right)^2
    +
    \\
    & \left(\frac{\partial H_y}{\partial t}(x^i,y^i,t^j) -
    \frac{1}{\mu}\frac{\partial E_z}{\partial x}(x^i,y^i,t^j)\right)^2
    \Bigg]
\end{aligned}
\label{eqn:2D_resLoss}
\end{equation}

Here, the permittivity and permeability are taken as $\varepsilon = \mu = 1$, corresponding to vacuum.

The total loss function is constructed as a weighted sum of its components:

\begin{equation}
    \mathcal{L}(\mathbf{x}, t) = 
    50 \mathcal{L}_{\text{IC}} + 
    50 \mathcal{L}_{\text{BC}} + \mathcal{L}_{\text{res}}
    \label{eqn:2D_lossTot}
\end{equation}
\newline

To assess the contribution of each architectural component, an ablation study was conducted by systematically removing RFF, strict periodicity, and causal training. Each variant was evaluated using the relative $L_2$ error against the high-resolution reference solution, and the results are summarized in Table~\ref{tab:ablation_case2}. The complete model—including RFF, strict periodicity, and causality—achieved the lowest $L_2$ error. Removing causality training caused a moderate increase in error, while removing both causality and temporal periodicity led to a more pronounced degradation. Excluding RFF while keeping the other enhancements further increased the error, and using RFF alone without the other strategies resulted in the worst performance.

% ablation study
\begin{table}[h]
    \centering
    \caption{Ablation study for the case of pulse in a periodic box (Sec. \ref{sec:case2}).}
    \renewcommand{\arraystretch}{1.2} % Adjust row height for better readability
    \setlength{\tabcolsep}{4pt} % Adjust column spacing
    \begin{tabular}{clccccc}
        \toprule
        \textbf{\#} & \textbf{RFF} & \multicolumn{2}{c}{\textbf{Strict periodicity}} & \textbf{Causality training} & \multicolumn{2}{c}{\textbf{$L_2$ Error}} \\
        \cmidrule(lr){3-4} \cmidrule(lr){6-7}
        & & \textbf{in space ($x,y$)} & \textbf{in time ($t$)} & & \textbf{HL dim. = 64} & \textbf{HL dim. = 128} \\
        \midrule
        1 & \cmark & \cmark & \cmark & \cmark & \texttt{0.08471} & \texttt{0.04218} \\
        2 & \cmark & \cmark & \cmark & \xmark & \texttt{0.07853} & \texttt{0.04272} \\
        3 & \cmark & \cmark & \xmark & \xmark & \texttt{0.11841} & \texttt{0.04895} \\
        4 & \xmark & \cmark & \cmark & \cmark & \texttt{0.19232} & \texttt{0.09558} \\
        5 & \cmark & \xmark & \xmark & \xmark & \texttt{0.51355} & \texttt{0.49885} \\
        \bottomrule
    \end{tabular}
    \label{tab:ablation_case2}
\end{table}

\begin{figure}[H]
    \centering
    \includegraphics[width=\linewidth]{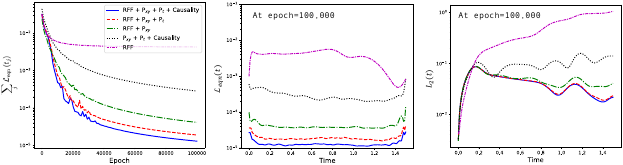}
    \caption{Total equation loss versus epoch (left), equation loss versus time after 100,000 epochs (middle), and $L_2$ error versus time after 100,000 epochs (right). Plots for the different architectural choices are presented. A hidden layer dimension of 128 was used.}
    \label{fig:case2_losses}
\end{figure}

Figure~\ref{fig:case2_losses} provides a comprehensive view of how each model variant evolves during training and how errors are distributed across time. The left plot shows the total residual loss as a function of training epochs. Models with full architecture—incorporating RFF, strict periodicity, and temporal causality—consistently outperform ablated variants in terms of convergence speed and final loss value. The full model achieves the lowest residual loss, though the difference compared to the model without causality is modest, suggesting that causality mainly improves temporal stability rather than raw convergence rate. Notably, the model using RFF alone converges significantly slower and reaches a much higher final loss, indicating that RFF alone is insufficient to regularize learning in this setting. This interpretation is supported by the middle plot, which shows the residual loss across time after 100,000 training epochs. The model with causality shows slightly more uniform loss across the time domain, whereas the model without causality has slightly elevated residuals at later time steps. The right plot displays the time-dependent $L_2$ error. As expected, all models start with low error at $t = 0$, due to supervision from the initial condition. The full model maintains a slightly lower error throughout the time domain, but again, the difference between the full model and the one without causality is relatively small. 

\begin{figure}[H]
    \centering
    \includegraphics[width=1.0\textwidth]{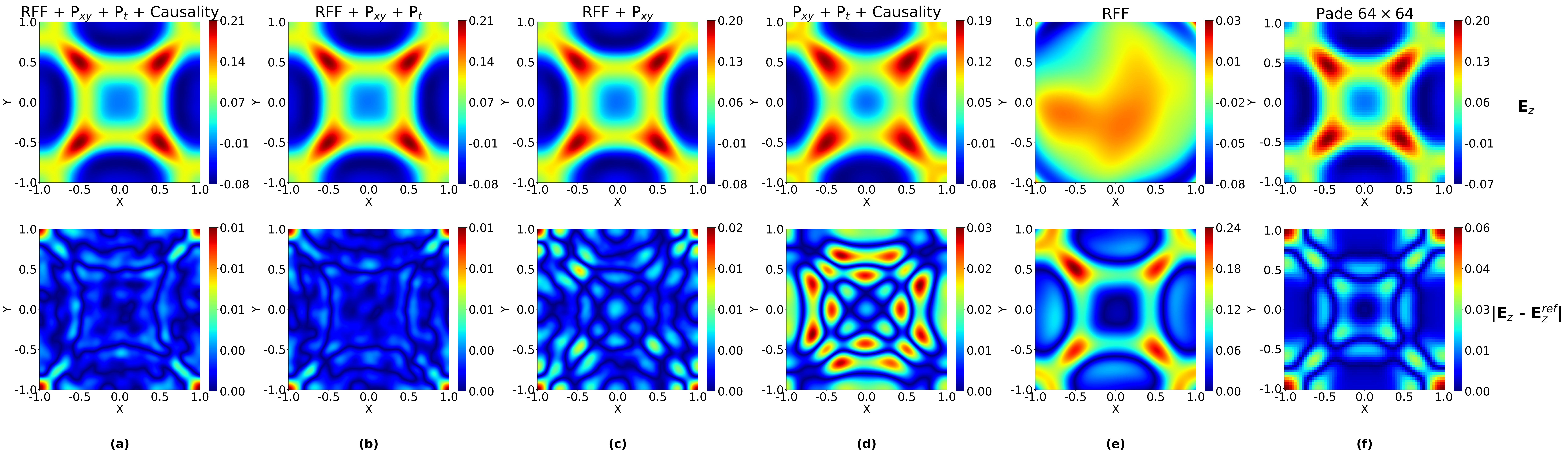}
    \caption{Electric field $E_z$ distribution (top row) and the corresponding absolute errors (bottom row) at $t = 1.5$ for the pulse-in-a-box test case, computed using (a–e) various PINN configurations and (f) the Pade numerical scheme. The Pade scheme was used to generate a reference solution on a $505 \times 505$ spatial grid. For comparison, subplot (f) presents the discrepancy between Pade solutions computed on $64 \times 64$ and $505 \times 505$ grids.}
\label{fig:Ez_comparison_testCase_2}
\end{figure}

Figure~\ref{fig:Ez_comparison_testCase_2} further illustrates the impact of these components on spatial field accuracy. Sub-figures (a)–(e) show qualitative comparisons between PINN-predicted fields (top rows) and their corresponding absolute errors (bottom rows), sorted from best to worst based on ablation configuration. Comparing (a) and (b), the solutions of the full model and the no-causality model are nearly indistinguishable by eye, and their absolute error maps show only slight differences in magnitude and structure. This reinforces that causality, while beneficial, is not strictly necessary for high accuracy in this problem when RFF and periodicity are already applied. By contrast, models (c–d) show clear deterioration as more components are removed, with increasing spatial error. The worst case (e), using only RFF, exhibits strong artifacts and widespread error, with distortion of field patterns, and breakdown of symmetry. For comparison, sub-figure (f) shows the absolute error of a low-resolution Pade scheme, which performs similarly to the least effective PINN model.

\subsection{2-D Gaussian pulse interaction with a dielectric medium} \label{sec:case3}
This test case involves a 2-D Gaussian electric pulse evolving within a non-homogeneous medium that contains a dielectric region. Unlike previous cases, the presence of spatially varying material properties introduces asymmetry in the solution and violates periodicity, making this a more realistic and challenging scenario. The computational domain is defined over \( x,y \in [-1,1] \), and the simulation time spans \( t \in [0,0.7] \). Collocation points are uniformly distributed with resolutions \( N_x = 64 \), \( N_y = 64 \), and \( N_t = 64 \). Although the problem setup is not strictly periodic in space, we exploit the fact that the pulse remains well-contained within the domain over the simulated time, which enables the use of periodic boundary conditions for practical purposes, even though the problem itself lacks strict periodicity. 

The domain is divided into two regions with distinct permittivity values:

\begin{itemize}
    \item For \( x \in [-1.0, 0.4] \): vacuum with \( \epsilon = 1 \).
    \item For \( x \in (0.4, 1.0] \): dielectric region with \( \epsilon = 4 \).
\end{itemize}

The initial condition is identical in form to the second test case (a centered 2-D Gaussian in $E_{z}$, and zero magnetic fields), with the initial condition loss and residual loss defined as in Eq.~\ref{eqn:pulse_testcase2_Ez}-\ref{eqn:pulse_testcase2_Hy}. The only difference lies in the spatially varying permittivity used in the residual term. The total loss function follows the structure of Eq.~\ref{eqn:2D_lossTot}, with boundary loss included or excluded depending on the model variant. 

An ablation study was conducted to evaluate the contributions of RFF, spatial and temporal periodicity, and causality training. The results are summarized in Table~\ref{tab:ablation_case3}. Contrary to earlier observations, the full model including all enhancements did not yield the lowest $L_2$ error. In fact, removing spatial periodicity consistently improved performance, suggesting that enforcing spatial periodicity in a medium with sharp material transitions (i.e., a dielectric interface) introduces artificial constraints that hinder generalization. The best-performing configuration excluded spatial periodicity while retaining RFF, temporal periodicity, and causality (row 2), achieving the lowest $L_2$ error across both network sizes. Removing causality alone had a minimal impact on error (comparing rows 1 and 4). Meanwhile, the worst performance was observed when both RFF and causality were excluded (row 7), confirming that RFF remains an essential component for resolving fine-scale features in nonhomogeneous domains.

\begin{figure}[H]
    \centering
    \includegraphics[width=\linewidth]{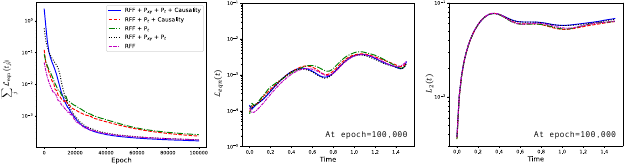}
    \caption{Total equation loss versus epoch (left), equation loss versus time after 100,000 epochs (middle), and $L_2$ error versus time after 100,000 epochs (right). Plots for the different architectural choices are presented.}
    \label{fig:case3_losses}
\end{figure}

Figure~\ref{fig:case3_losses} illustrates training dynamics across these configurations. The leftmost plot shows the total residual loss versus epoch, with all models eventually converging. However, the magnitude and slope of convergence are not strongly indicative of solution quality, particularly in complex spatial domains. The center plot shows the residual loss as a function of time at the end of training, with modest differences among models. The rightmost plot presents the time-dependent 
$L_2 (t)$ error. As expected, all models exhibit minimal error at $t = 0$, but diverge as the solution evolves. Notably, models without spatial periodicity, especially those retaining RFF, maintain lower errors throughout the time domain, reflecting their ability to flexibly adapt to asymmetric material interfaces.

% ablation study
\begin{table}[h]
    \centering
    \caption{Ablation study for the case of interaction of 2D pulse with a dielectric medium. (Sec. \ref{sec:case3})}
    \renewcommand{\arraystretch}{1.2} % Adjust row height for better readability
    \setlength{\tabcolsep}{4pt} % Adjust column spacing
    \begin{tabular}{clccccc}
        \toprule
        \textbf{\#} & \textbf{RFF} & \multicolumn{2}{c}{\textbf{Strict periodicity}} & \textbf{Causality training} & \multicolumn{2}{c}{\textbf{$L_2$ Error}} \\
        \cmidrule(lr){3-4} \cmidrule(lr){6-7}
        & & \textbf{in space ($x,y$)} & \textbf{in time ($t$)} & & \textbf{HL dim. = 64} & \textbf{HL dim. = 128} \\
        \midrule
        1 & \cmark & \cmark & \cmark & \cmark & \texttt{0.06383}  & \texttt{0.05940} \\
        2 & \cmark & \xmark & \cmark & \cmark & \texttt{0.05920}  & \texttt{0.05670} \\
        3 & \cmark & \xmark & \cmark & \xmark & \texttt{0.06953}  & \texttt{0.05696} \\
        4 & \cmark & \cmark & \cmark & \xmark & \texttt{0.06295}  & \texttt{0.05905} \\
        5 & \cmark & \xmark & \xmark & \xmark & \texttt{0.05930}  & \texttt{0.05753} \\
        6 & \cmark & \cmark & \xmark & \xmark & \texttt{0.07666}  & \texttt{0.06428} \\
        7 & \xmark & \cmark & \cmark & \xmark & \texttt{0.12211}  & \texttt{0.08173} \\
        \bottomrule
    \end{tabular}
    \label{tab:ablation_case3}
\end{table}

The full model (a) produces a reasonable prediction, but exhibits elevated error near the domain boundaries rather than at the dielectric interface. Configuration (b), which omits spatial periodicity while retaining RFF, $P_{t}$, and causality, yields improved accuracy both globally and near the boundaries, highlighting that enforcing spatial periodicity in a non-periodic domain can introduce error. As more architectural constraints are removed in (c)–(e), solution quality gradually degrades, though the RFF-only model (e) remains competitive. Subfigure (f) compares the low-resolution Pade solution against the high-resolution reference, serving as a numerical baseline. The error magnitude and structure are similar to those produced by the least-constrained PINN models, confirming that even basic PINN configurations can rival traditional solvers at lower resolution—while avoiding grid generation and post-processing steps.

\begin{figure}[H]
    \centering
    \includegraphics[width=1.0\textwidth]{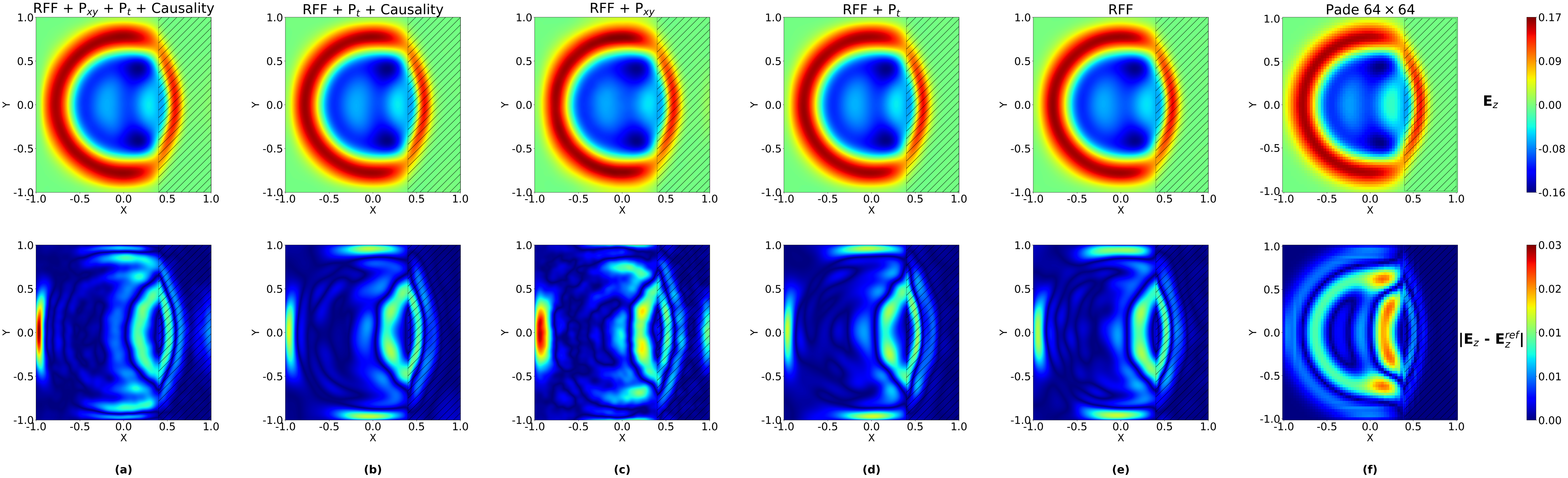}
    \caption{Electric field $E_z$ distribution (top row) and corresponding absolute errors (bottom row) at $t = 1.5$ for the 2-D pulse interacting with a dielectric medium. Subplots (a)-(e) shows results from different PINN configurations, each compared against a high-resolution Pade reference solution computed on $505 \times 505$ grid. Subplot (f) presents the discrepancy between Pade solutions computed on $64 \times 64$ and $505 \times 505$ grids.}
    \label{fig:Ez_comparison_testCase_3}
\end{figure}

\section{Spatio-temporal Convergence Rate Using Empirical Neural Tangent Kernel}
One of the main challenges in training PINNs is their inefficient convergence when minimizing the residual loss. Analyzing the convergence rate as a function of space ($\mathbf{x}$) and time ($t$) provides valuable insight into regions where the network is effectively learning versus those where it is not. To estimate the local convergence rate across space-time, we leverage Neural Tangent Kernel (NTK) theory \cite{jacot2018neural}. While NTK theory strictly applies to infinitely wide networks trained with infinitesimal learning rates, it has also been shown to offer meaningful approximations for finite-width networks \cite{arora2019exact, lee2019wide, yang2019scaling}. According to the theory, the convergence rate is proportional to the sum of the eigenvalues of the NTK matrix. So, we compute the NTK matrix using a specific set of collocation points sampled from a localized region in the $\mathbf{x}$-$t$ domain and compute the eigenvalue sum. Given a PINN model \( f(\mathbf{x}, \boldsymbol{\theta}) \) parameterized by \( \boldsymbol{\theta} \in \mathbb{R}^P \),and governed by a physics-based residual  \( \mathcal{R} \), the NTK can be defined as:

\begin{equation}
    \Theta(\mathbf{x}, t) = 
    \underbrace{\nabla_{\boldsymbol{\theta}} \mathcal{R}(\mathbf{x}, t, \boldsymbol{\theta})^\top}_{N \times P}
    \underbrace{\nabla_{\boldsymbol{\theta}} \mathcal{R}(\mathbf{x}, t, \boldsymbol{\theta})}_{P \times N}
    \in \mathbb{R}^{N \times N},
\end{equation}

where $N$ is the number of collocation points used in evaluating the NTK corresponding to the specific spatio-temporal sub-domain and $P$ indicates the total number of network parameters. The NTK based spatio-temporal convergence rate can be thus defined as:

\begin{equation}
    C(\mathbf{x}, t)=\frac{\sum_{k=1}^{N} \lambda_k(\mathbf{x}, t)}{N}
\end{equation}

which refers to the mean of the eigenvalues \( \lambda \) of the NTK matrix evaluated at a specific set of spatio-temporal points, and serves as an indicator of the \textit{local convergence rate} at those specific set of collocation points in space-time. 

% In their calculation, the network is provided with all spatial coordinates \( \mathbf{x} \in \mathbb{R}^{N_{xy} \times 1} \) for a specific time instance \( t \in \mathbb{R}^{N_{xy} \times 1} \) to compute the convergence rate value for that time instance. By repeating this process for multiple time values, the NTK based convergence rate $C(t)$ can be obtained, which indicates the degree of convergence the network is able at each time.

\begin{figure}[H]
    \centering
    \includegraphics[width=\linewidth]{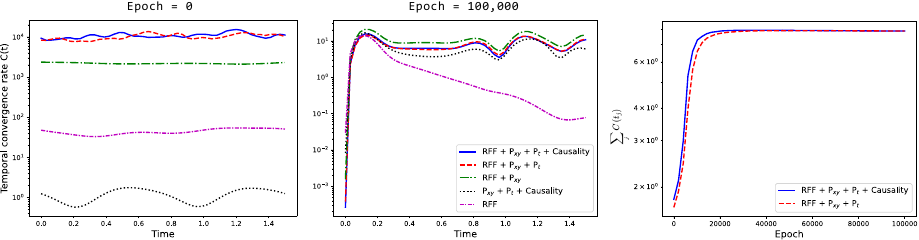}
    \caption{Temporal convergence rate $C(t)$ for different PINN configurations is shown before the (a) first epoch and (b) last epoch. The convergence rate is compared at different epochs for PINN configurations with and without temporally causal training. A close agreement is observed between the two configurations.}
    \label{fig:case2_ntk}
\end{figure}

As demonstrated by Wang et al.~\cite{wang2022}, the NTK-based convergence rate can also be evaluated purely as a function of time by computing the NTK across the spatial domain at fixed temporal instances. Following this approach, we analyze the time-dependent convergence rate for pulse-in-a-box case (Section~\ref{sec:case2}). Figure~\ref{fig:case2_ntk} presents the NTK-derived convergence rates at two key stages of training: epoch 0 (random initialization) and epoch 100,000 (end of training). At initialization, clear differences emerge between the various network configurations. In particular, networks incorporating both RFF and spatio-temporal periodicity (shown in blue and red) exhibit substantially higher convergence rates compared to the baseline. This behavior is consistent with prior findings~\cite{wang2020eigenvector}, which show that Fourier features mitigate spectral bias and improve the learning of high-frequency content. Moreover, the inclusion of strict periodicity constraints in both space and time further boosts the initial convergence rate. Throughout training, architectures that combine RFF with spatio-temporal periodicity consistently maintain superior convergence performance. Notably, the convergence rate trends at epoch 0 correlate well with the final $L_2$ errors reported in Table~\ref{tab:ablation_case2}, underscoring the predictive value of NTK-based diagnostics for assessing network trainability. As training progresses, the convergence rate drops sharply near \( t = 0 \), where the network is explicitly supervised to match the initial condition. Beyond this initial phase, \( C(t) \) stabilizes around a value of 10 with modest fluctuations. Notably, the temporal profile of \( C(t) \) closely follows that of the time-dependent $L_2$ error, suggesting a strong correlation between them. This observation motivates further investigation into whether a similar relationship exists in the spatial domain.

\begin{figure}[H]
    \centering
    \includegraphics[width=\linewidth]{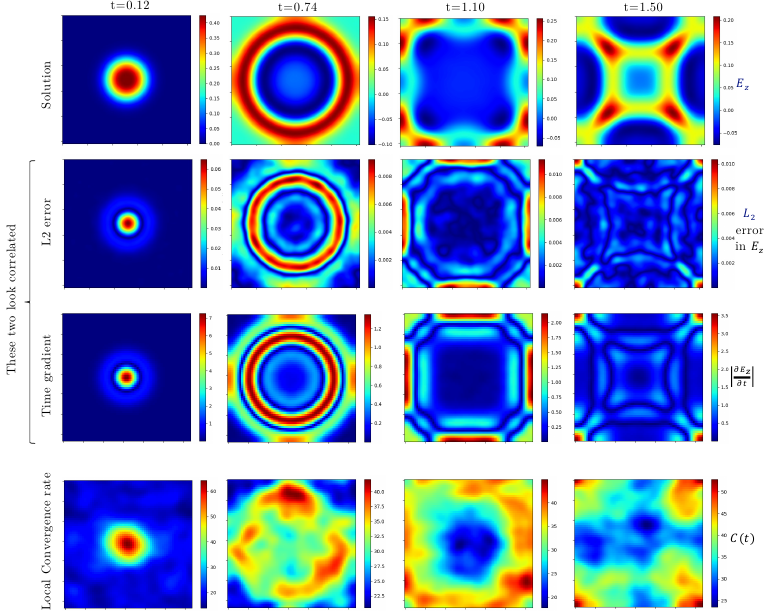}
    \caption{Instantaneous $E_z$ field (row 1), $L_2$ error (row 2), auto-differentiated time derivative $\frac{\partial E_z}{\partial t}$ (row 3), and NTK-based convergence rate (row 4) at different time instants for the pulse-in-a-box case (Sec.~\ref{sec:case2}). Notably, the $L_2$ error aligns with regions of high $\frac{\partial E_z}{\partial t}$, indicating strong correlation. All plots correspond to configuration \#1 in Table~\ref{tab:ablation_case2}, which uses RFF + P$_{xy}$ + P$_t$ + Causality.}
    \label{fig:ntk}
\end{figure}

Figure~\ref{fig:ntk} presents a detailed comparison at four representative time instances for the pulse-in-a-box case (Section~\ref{sec:case3}). Each column corresponds to a fixed time snapshot and shows: the instantaneous $E_z$ solution (row 1), corresponding relative $L_2$ error (row 2), automatic-differentiation-based time derivative of $E_z$ (row 3), and spatio-temporal NTK-based convergence rate (row 4). A strong visual correlation is observed between the error map (row 2) and the time derivative (row 3), indicating that regions with steep temporal gradients tend to exhibit high prediction error. This suggests that the network struggles to fit solution regions undergoing rapid temporal evolution. To assess whether this is a result of limited temporal sampling, we increased the number of time collocation points by a factor of two. However, the correlation between high $L_2$ error and steep time derivatives persisted, indicating that this behavior is likely rooted in the inherent limitations of the training dynamics rather than insufficient temporal resolution. The NTK-based convergence rate maps (row 4) further highlight this limitation. While some regions exhibit fast convergence (red) and others slow convergence (blue), their spatial distribution does not clearly align with the high-error regions identified in row 2. In fact, the convergence rate appears irregular and uncorrelated with both the prediction error and the physical dynamics of the problem. This indicates that the network’s optimization does not inherently focus on error-prone regions, resulting in uneven learning across the domain. These findings reinforce that PINN training does not necessarily direct learning effort toward regions of greatest difficulty. Instead, the spatio-temporal convergence landscape is shaped by complex interactions between the network architecture, loss formulation, and optimization dynamics. Identifying and addressing this inefficiency is essential for improving the robustness and reliability of PINN-based solvers in stiff or rapidly evolving problems.

\section{Conclusions}
In this study, we evaluated the effectiveness of PINNs for solving unsteady Maxwell’s equations and compared their performance against two conventional numerical methods: the FDTD method and a high-order compact Pade scheme with filtering. Through three representative test cases—including a 1-D Gaussian pulse in free space and two 2-D scenarios involving periodic and dielectric media—we systematically analyzed the influence of key architectural features and training strategies on solution accuracy and convergence behavior:

\begin{itemize}
    \item In the 1-D case, PINNs accurately resolved sharp discontinuities without introducing oscillations, outperforming both FDTD and Pade methods near steep gradients. This demonstrates the advantage of the mesh-free PINN formulation in resolving transient electromagnetic fields, especially in smooth domains with localized dynamics.
    \item For the 2-D periodic case, an ablation study was conducted to assess the effects of RFF, strict periodicity enforcement, and causality training. The full model incorporating all enhancements achieved the lowest $L_2$ error, with RFF and periodicity contributing most significantly to performance. Causality training had a smaller but stabilizing effect, particularly in reducing temporal drift.
    \item In the more complex 2-D dielectric case, the benefits of architectural enhancements became more nuanced. Multiple configurations achieved comparable accuracy, with RFF being the most consistently beneficial. Enforcing spatial periodicity degraded performance due to material asymmetry, while causality and temporal periodicity had only marginal effects.
\end{itemize}

To further understand convergence dynamics, we applied a NTK-based diagnostic to assess local learning behavior across space and time. While convergence rate aligned closely with $L_2 (t)$ error, no such correlation was observed spatially. Regions with steep temporal gradients consistently exhibited higher errors and slower convergence, even when time resolution was increased, indicating an intrinsic limitation of current PINN training dynamics. The network fails to prioritize error-prone regions, which contributes to uneven learning and residual hotspots.

In summary, this work demonstrates that PINNs, when equipped with appropriately chosen enhancements such as RFF and domain-aware priors, can match or exceed traditional solvers for Maxwell’s equations in structured and smooth domains. However, challenges remain in handling spatial inhomogeneity and adaptively directing learning to regions of difficulty. Future research should focus on dynamic sampling, adaptive loss weighting, and more problem-aware architectures to improve PINN robustness and scalability for real-world electromagnetic applications.

% Overall, we conclude that strict periodicity and RFF are the main components that help with the accuracy of the solution and aid the convergence of the network when the electromagnetic wave reaches the domain of the boundaries. 

% type here
% key takeaways:
% 1. PINNs struggle to minimize errors near the regions of large gradients.
% 2. When the PINNs is already equipped with enough capabilities to converge to a solution, temporal cuasality aware training does not effect the convergence rate evolution significantly. We obtain to this inference based on the fact that the L2 errors and 
% 3. 

\appendix

\section{Discretization of Maxwell's equations} \label{app:numerics}
In Section \ref{sec:expts}, we have compared the PINNs solution with results from two widely used numerical methods: (a) the FDTD method and (b) a high-order Pade scheme with filtering. Below, we briefly outline the numerical algorithms for both approaches, using the 2-D Maxwell’s equations (Eq. \ref{eq:maxwell_2d_1}–\ref{eq:maxwell_2d_3}) as the governing framework.

\subsection{Finite difference time domain (FDTD)}

The FDTD method is a widely used numerical scheme for solving Maxwell’s equations \cite{yee1966numerical, taflove2005computational}. The FDTD implemented in the present work achieves second-order accuracy in both space and time. The method employs a staggered grid for spatial discretization using central differencing and utilizes a leapfrog scheme for time integration. Firstly, a uniform collocated grid is constructed with \( N_x \) and \( N_y \) points along the \( x \)- and \( y \)-directions, respectively. The spatial discretization results in a resolution of \( [\Delta x, \Delta y] = \left[\frac{L_x}{N_x-1}, \frac{L_y}{N_y-1}\right] \), where \( L_x \) and \( L_y \) denote the domain dimensions. The time step \( \Delta t \) is chosen as \( \Delta t = 0.5 \min(\Delta x, \Delta y) \) to ensure numerical stability. The field component \( E_z \) is updated iteratively as:

\begin{equation} \label{eqn:EzUpdate}
    E_z^{n+1}(i,j) = E_z^{n}(i,j) + \frac{\Delta t}{\varepsilon} \left[ \left( \frac{\partial H_y}{\partial x} \right)_{i,j} - \left( \frac{\partial H_x}{\partial y} \right)_{i,j} \right],
\end{equation}

where the spatial derivatives of \( H_y \) and \( H_x \) are approximated as:

\begin{equation}
    \left( \frac{\partial H_y}{\partial x} \right)_{i,j} = \frac{H_y^{n+1/2}(i+1/2,j) - H_y^{n+1/2}(i-1/2,j)}{\Delta x},
\end{equation}
\begin{equation}
    \left( \frac{\partial H_x}{\partial y} \right)_{i,j} = \frac{H_x^{n+1/2}(i,j+1/2) - H_x^{n+1/2}(i,j-1/2)}{\Delta y}.
\end{equation}

Here, \( i \) and \( j \) denote the structured grid indices along the \( x \)- and \( y \)-directions, respectively. Since \( H_y^{n+1/2}(i+1/2,j) \) and \( H_x^{n+1/2}(i,j+1/2) \) values on the staggered grid are unknown, they are computed using the magnetic field update equations as follows:

\begin{equation}
    H_x^{n+1/2}(i,j+1/2) = H_x^{n-1/2}(i,j+1/2) - \frac{\Delta t}{\mu \Delta y} \left[ E_z^n(i,j+1) - E_z^n(i,j) \right],
\end{equation}

\begin{equation}
    H_y^{n+1/2}(i+1/2,j) = H_y^{n-1/2}(i+1/2,j) + \frac{\Delta t}{\mu \Delta x} \left[ E_z^n(i+1,j) - E_z^n(i,j) \right].
\end{equation}

% Boundaries  
It is worth noting that $E_x$ is updated at integer time steps ($n+1, n+1, ...$), while $H_x$ and $H_y$ are updated at half time steps ($n+\frac{1}{2},n+\frac{3}{2},...$) staggered in time. The values at the boundary points are not updated iteratively but are instead specified explicitly. For Dirichlet boundary conditions, the solution at the boundary is set to a fixed value as defined in the problem. To impose Neumann boundary conditions, a simple approach is used, where the boundary value is set equal to the nearest interior point. This effectively enforces a weak zero-gradient condition at the boundary.  

\subsection{High-order Pade scheme with filtering}
For the high-order Pade scheme, a fourth-order accurate discretization in space \cite{lele1992compact, visbal2002use, kakumani2022use} is employed. The algorithm follows the same structure as the one described for the FDTD method, except that the derivatives \( \frac{\partial H_y}{\partial x} \), \( \frac{\partial H_x}{\partial y} \), \( \frac{\partial E_z}{\partial y} \), and \( \frac{\partial E_z}{\partial x} \) are computed using the following implicit formula:

\begin{equation}
\frac{1}{4} \phi_{i-1}^{\prime}+\phi_i^{\prime}+ \frac{1}{4}  \phi_{i+1}^{\prime}= \frac{2}{3 \Delta x} \left(\phi_{i+1}-\phi_{i-1}\right) 
\end{equation}

Since this formulation is implicit in nature, a matrix inversion is required to obtain the gradient values at each spatial location. To suppress the non-physical oscillations that arise from the dispersion error of the central scheme, the field variables $E_z$, $H_x$, and $H_y$ are filtered through a 6th order low pass filter:

\begin{equation}
\alpha_f \hat{\phi}_{i-1}+ \hat{\phi}_i+ \alpha_f \hat{\phi}_{i+1}=\Sigma a_n\left(\phi_{i+n}+\phi_{i-n}\right) / 2, \quad n=0,1,2,3.
\end{equation}

Where \( \phi \) and \( \hat{\phi} \) represent the pre- and post-filtered field values, respectively. The filter coefficient \( \alpha_f \) is set to \( 0.48 \). The coefficients \( a_0 \), \( a_1 \), and \( a_2 \) are given by:

\begin{equation}
    a_0 = \frac{5}{8} + \frac{3 \alpha_f}{4}, \quad a_1 = \frac{1}{4} + \frac{\alpha_f}{2}, \quad a_2 = \frac{-1}{16} + \frac{\alpha_f}{8}.
\end{equation}

One-sided and partially one-sided formulations are employed for the derivative and filtering calculations at the boundaries, ensuring the same order of accuracy is maintained. For a detailed description of the boundary formulations, readers are referred to \cite{visbal2002use}.

\bibliographystyle{unsrt}
\bibliography{references}

\end{document}